\documentclass[preprint]{aastex}
\usepackage{graphicx}
\usepackage{lscape}
\usepackage{subfigure}

\input{epsf}

\newcommand{\x}{\mbox{${}\times{}$}}
\def\lsim {$\rlap{\raise.4ex\hbox{$<$}}\lower.55ex\hbox{$\sim$}\,$}

\begin{document}
               

\title {\bf Optical Observations of the Transiting Exoplanet GJ 1214b}
\author{Johanna K. Teske\altaffilmark{1,2}, Jake
  D. Turner\altaffilmark{1,2}, Matthias Mueller\altaffilmark{3}, Caitlin A. Griffith\altaffilmark{2}}

\altaffiltext{1}{Steward Observatory, University of Arizona, Tucson, AZ, 85721, USA \\ 
email: jkteske@as.arizona.edu}
\altaffiltext{2}{Lunar and Planetary Laboratory, University of
  Arizona, Tucson, AZ, 85721, USA}
\altaffiltext{3}{Leibniz-Institut für Astrophysik Potsdam (AIP), An
  der Sternwarte 16, D-14482 Potsdam, Germany}

 
\begin{abstract}

We observed nine primary transits of the super-Earth exoplanet GJ
1214b in several optical photometric bands from March to August 2012,
with the goal of constraining the short-wavelength slope of the
spectrum of GJ 1214b. Our observations were conducted on the Kuiper
1.55 m telescope in Arizona and the STELLA-I robotic 1.2 m telescope in
Tenerife, Spain. From the derived light curves we extracted transit
depths in $R$ (0.65 $\mu$m), $V$ (0.55 $\mu$m), and $g'$ (0.475
$\mu$m) bands. Most previous observations of this
exoplanet suggest a flat spectrum varying little with wavelength from
the near-infrared to the optical, corresponding to a low-scale-height,
high-molecular-weight atmosphere. However, a handful of observations
around $K_{s}$ band ($\sim$2.15 $\mu$m) and $g$-band ($\sim$0.46
$\mu$m) are inconsistent with this scenario and suggest a variation on
a hydrogen- or water-dominated atmosphere that also contains a haze
layer of small particles. In particular, the $g$-band observations of de Mooij et
al.\,(2012), consistent with Rayleigh scattering, limit the
potential atmosphere compositions of GJ 1214b due to the increasing
slope at optical wavelengths (Howe \& Burrows 2012). We find that our results overlap
within errors the short-wavelength observations of de Mooij et
al.\,(2012), but are also consistent with a spectral
slope of zero in GJ 1214b in the optical wavelength region. Our observations thus allow for a larger suite of possible atmosphere compositions, including those with a high-molecular-weight and/or hazes.
\end{abstract}


\section{Introduction}

Since the detection of the `super-Earth' transiting extrasolar planet GJ 1214b (Charbonneau et al.\,2009), its composition has been a topic of interest and debate. Discovered by the MEarth program, GJ 1214b has a radius (2.85$\pm$0.20 R$_{\earth}$; Harps{\o}e et al.\,2013) and mass (6.26$\pm$0.86 M$_{\earth}$; Harps{\o}e et al.\,2013) only slightly larger that of the Earth, and transits a near by (13 pc) M star (0.216 $\pm$ 0.012 R$_{\sun}$; Harps{\o}e et al.\,2013) with an orbital period of 1.5804 days and a semi-major axis of 0.0197 AU (Harps{\o}e et al.\,2013). This causes a planet-to-star flux ratio comparable to that of a Jupiter-sized planet orbiting the Sun, and makes it one of only a handful of `super-Earth' atmospheres that currently can be investigated with transit spectroscopy (Charbonneau et al.\,2009). GJ 1214b represents a unique opportunity to study a planetary object unlike those in our Solar System, yet potentially similar to a large fraction of currently-detected exoplanets, many of which are smaller than Jupiter-sized (Borucki et al.\,2012; Muirhead et al.\,2012; Borucki et al.\,2010; Howard et al.\,2010).

The mass and radius of GJ 1214b imply a low density of 1.87$\pm$0.40 g
cm$^{-3}$ ($\sim$0.35$\rho_{\rm{Earth}}$; Rogers \& Seager 2010) and
suggest that GJ 1214b cannot be composed of rock and water ice alone,
but likely has a significant gaseous atmosphere (Bean, Miller-Ricci Kempton \& Homeier 2010; Miller-Ricci \&
Fortney\,2010; Kundurthy et al.\,2011). Models of its interior structure indicate GJ 1214b's composition is most likely either (i) a mini-Neptune made of mainly solid rock and ice with a significant hydrogen-dominated atmosphere accreted from its protoplanetary nebula, (ii) a world composed mainly of water ice with a secondary water vapor envelope formed by sublimination, or (iii) an object composed of purely rocky material with a hydrogen-dominated atmosphere formed by outgassing (Rogers \& Seager 2010). A determination of the current composition of GJ 1214b will shed light on this planet's formation history, and thus potentially that of other super-Earth planets. If GJ 1214b's atmosphere is largely hydrogen, it likely formed from the accretion of proto-solar nebular gas or from the outgassing of significant amounts of hydrogen during the planet's cooling and solidification (Miller-Ricci Kempton, Zahnle \& Fortney\,2012). However, if instead the atmosphere is water-rich, then GJ 1214b could have formed from ice-rich material farther out in the protoplanetary disk before migrating inwards towards the star. Alternatively or in addition, the planet could have accreted less hydrogen-dominated nebular gas in the first place, or lost by atmospheric escape any hydrogen-rich gas that it did accrete (Rogers \& Seager 2010).  

Transmission photometry and spectroscopy indicate the bulk composition of GJ 1214b by measuring the attenuation of stellar light as it passes through the limb of the exoplanet's atmosphere. The modulation in the spectrum with wavelength increases with the atmosphere's scale height, which is inversely proportional to the atmosphere's molecular weight. The modulation in GJ 1214b's spectrum thus distinguishes between its possible compositions, since models (i) and (iii) (listed above) will have a large scale height and show prominent spectral features from absorption by molecular hydrogen, whereas model (ii) will have relatively small spectral features and scale height.

Several studies using transmission observations to determine GJ
1214b's atmospheric scale height and ; Fraine et al.\,2013composition have been
published. From the optical ($\sim$0.6 $\mu$m) through the
near-infrared (4.5 $\mu$m), most measurements indicate no significant
spectral modulation with wavelength in the atmosphere of GJ 1214b
(Bean et al.\,2010; Bean et al.\,2011; Crossfield, Barman \& Hansen\,2011; D{\'e}sert et
al.\,2011; Berta et al.\,2012; Narita et al.\,2012). Taken together, these observations
suggest that GJ 1214b has a small scale height, and favors model (ii)
above, in which the exoplanet's atmosphere is dominated by water
rather than hydrogen. However, there are hints of deviation (albeit
with less statistical significance) from the
flat-spectrum model in $K_{s}$-band (2.15 $\mu$m; Croll et al.\,2011;
de Mooij et al.\,2012), $g$-band (0.46 $\mu$m; de Mooij et al.\,2012),
and $R$-band (0.65 $\mu$m; Murgas et al.\,2012). Including these
latter observations requires modification of the water-world
explanation. Collectively, the observations may alternatively be
explained by a hydrogen-dominated atmosphere with an opacity source
causing the muted spectral features (Miller-Ricci Kempton et
al.\,2012; Howe \& Burrows 2012). The increase in the radius-ratio observed at short wavelengths is roughly consistent with Rayleigh scattering in an atmosphere with a relatively high scale height.

The goal of this paper is to constrain the transmission spectrum of GJ
1214b in the optical wavelength bands ($\lesssim$ 0.70$\mu$m) in order
to study the short-wavelength slope and the scattering regime in GJ
1214b's atmosphere. If the short-wavelength data are indicative of a Rayleigh scattering power law, this indicates a relatively high scale height atmosphere, and scattering particles that are much smaller than the wavelength of light. A shallower slope in the short-wavelength data would indicate a particle-size closer to the Mie scattering regime, $\sim$1 $\mu$m, or a small scale height atmosphere. 

In Section 2 we give an overview of our observations and data reduction procedures. We discuss our transit light curve analysis in Section 3 and the implications of our results in Section 4. 

\section{Observations and Data Reduction}

Our $R$- and $V$-band observations of the transit of GJ 1214b were
conducted between March and June 2012 at the Steward Observatory 1.55 meter Kuiper Telescope on
Mt. Bigelow near Tucson, Arizona using the Mont4k CCD. The Mont4k CCD
contains a 4096$^{2}$ pixel sensor with a field of view (FOV) of
9.7'$\times$9.7'. We used 3x3 binning to achieve a resolution of
0.43''/pixel, and a 3072$\x$1024 pixel subframe with a field-of-view (FOV) of 7.28'$\times$2.43' to shorten read-out time to roughly 10 seconds. Our observations were taken with the 
Harris V (473-686 nm; FWHM 88 nm), and Harris R (550-900 nm; FWHM 138 nm) 
photometric band filters, and we did not defocus the telescope (GJ 1214A is not bright enough to saturate the detector with our short integration times). To ensure accurate time-keeping, an on-board clock was automatically synchronized with GPS every few seconds throughout the observational period. Due to excellent autoguiding, there was no more than a 4.4 pixels ($\sim$1.9'') drift in the x position and 2.1 pixels ($\sim$0.9'') in the y position of GJ 1214A in all our data sets for the Kuiper 1.55 m telescope (with averages of 0.03'' in the x position and 0.06'' in the y position). Our Kuiper 1.55 m observations are summarized in Table \ref{tab1}.

All $g'$-band (401-550 nm; FWHM 153 nm) transit observations were
taken between May and August 2012 with STELLA-I, a fully robotic 1.2 m telescope in Tenerife, Spain (Strassmeier et al.\,2010). Its wide field imager WiFSIP hosts a 4096$^{2}$ 15-micrometre pixel back-illuminated CCD. It images a FOV of 22'$\times$22' with a scale of 0.322''/pixel. Because of a sufficiently high density of suitable comparison stars in the field of GJ 1214A we applied a CCD window of 2000$^{2}$ pixels, reducing the field of view to about 11'$\times$11'. We did not defocus the telescope because there was no danger of saturation of GJ 1214A due to its faintness at blue wavelengths. Each transit observation lasted $\sim$3 hours covering a sufficient amount of out-of-transit baseline before and after the rather short transit of $\sim$53 minutes. The robotic telescope took 98 exposures per run with 90 s exposure time and 20 s overhead resulting in a cadence of about 110 seconds. Our STELLA-I 1.2 m observations are summarized in Table \ref{tab1}.

Using standard IRAF\footnote{IRAF is distributed by the National
  Optical Astronomy Observatory, which is operated by the Association
  of Universities for Research in Astronomy, Inc., under cooperative
  agreement with the National Science Foundation.} reduction
procedures, each of our Kuiper 1.55 m images were bias-subtracted and
flat-fielded. Turner et al.\,(2013) determined that using
different numbers of flat-field images (flats) in the reduction of
Kuiper Telescope/Monk4k data did not significantly reduce the noise in
the resulting images. Thus, to save time, we used 10 flats in all
sequential observations and reductions, as well as 10 bias frames taken during each observing run.

\begin{figure}[ht!]
\figurenum{1.1}
   \subfigure{\includegraphics[width=0.5\textwidth]{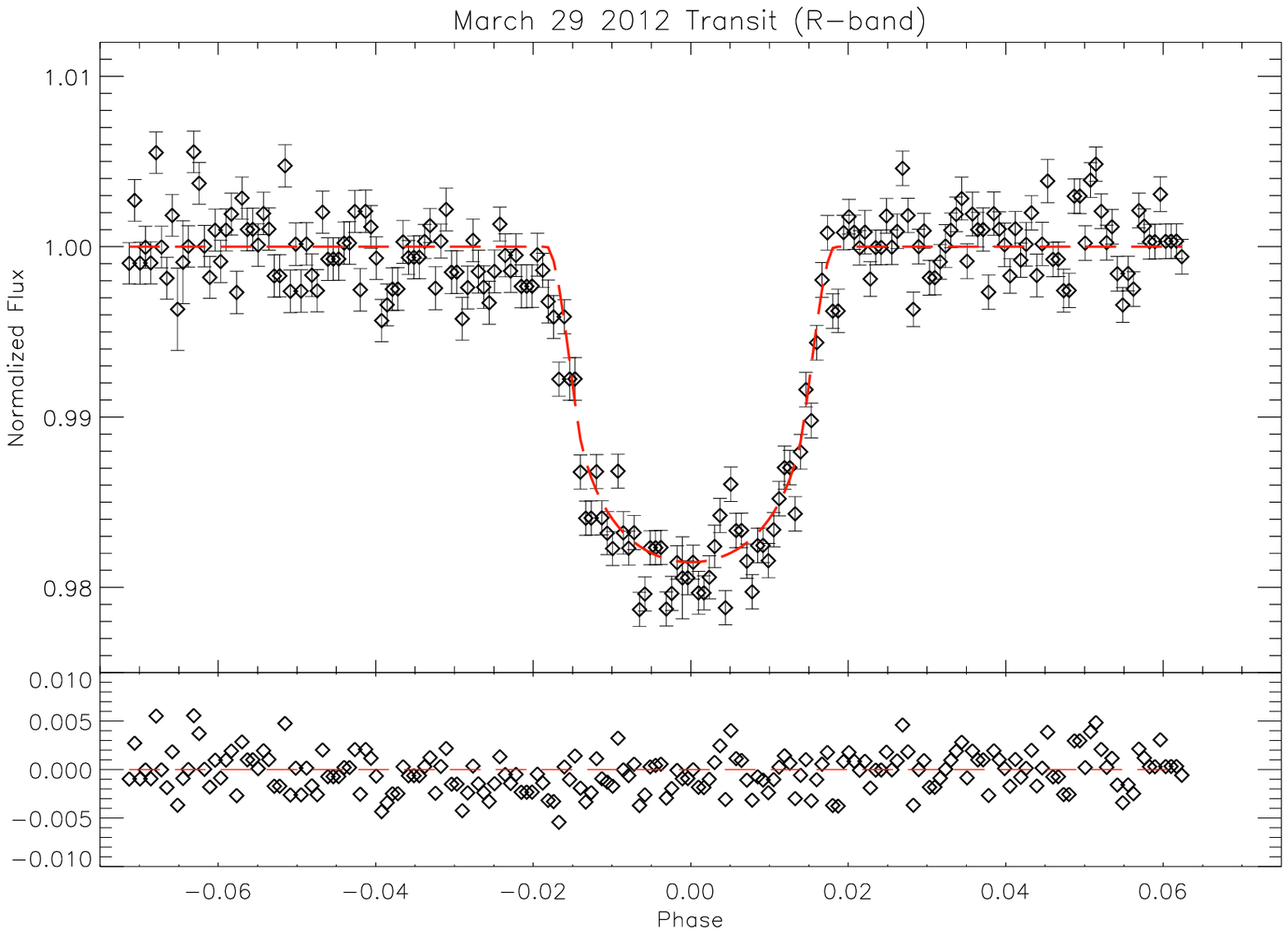}}
\quad
   \subfigure{\includegraphics[width=0.5\textwidth]{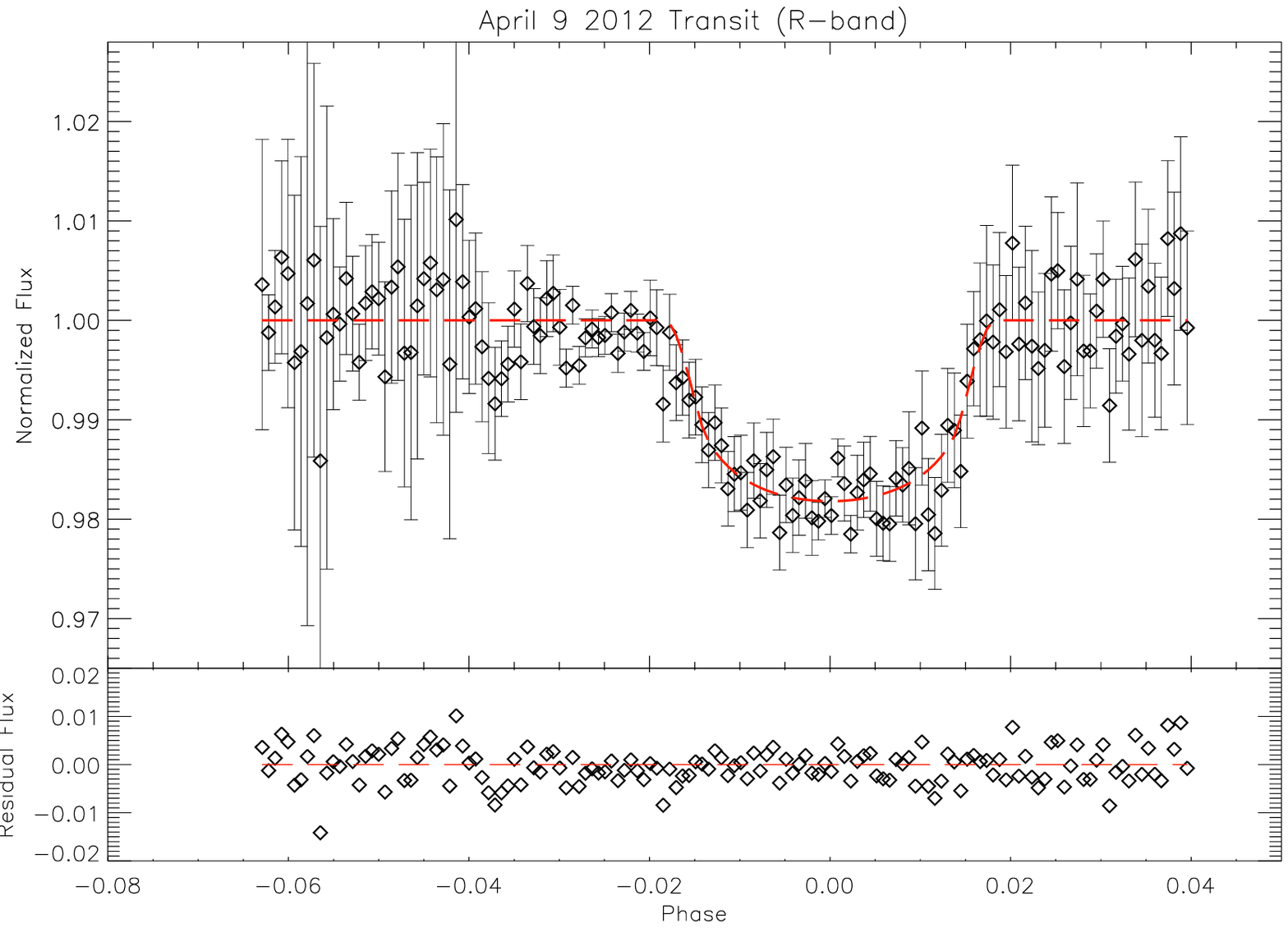}}

   \subfigure{\includegraphics[width=0.5\textwidth]{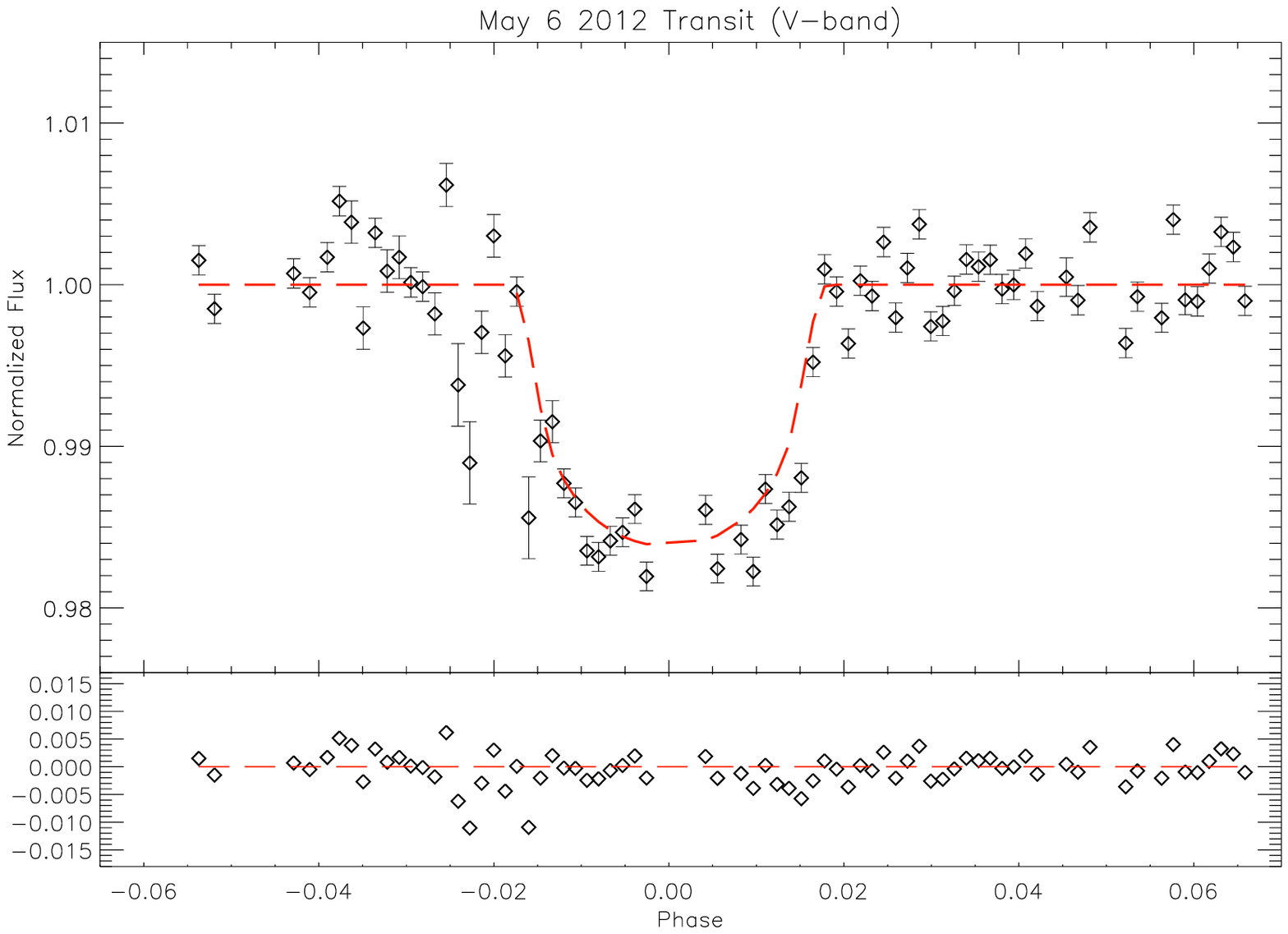}}
\quad
   \subfigure{\includegraphics[width=0.5\textwidth]{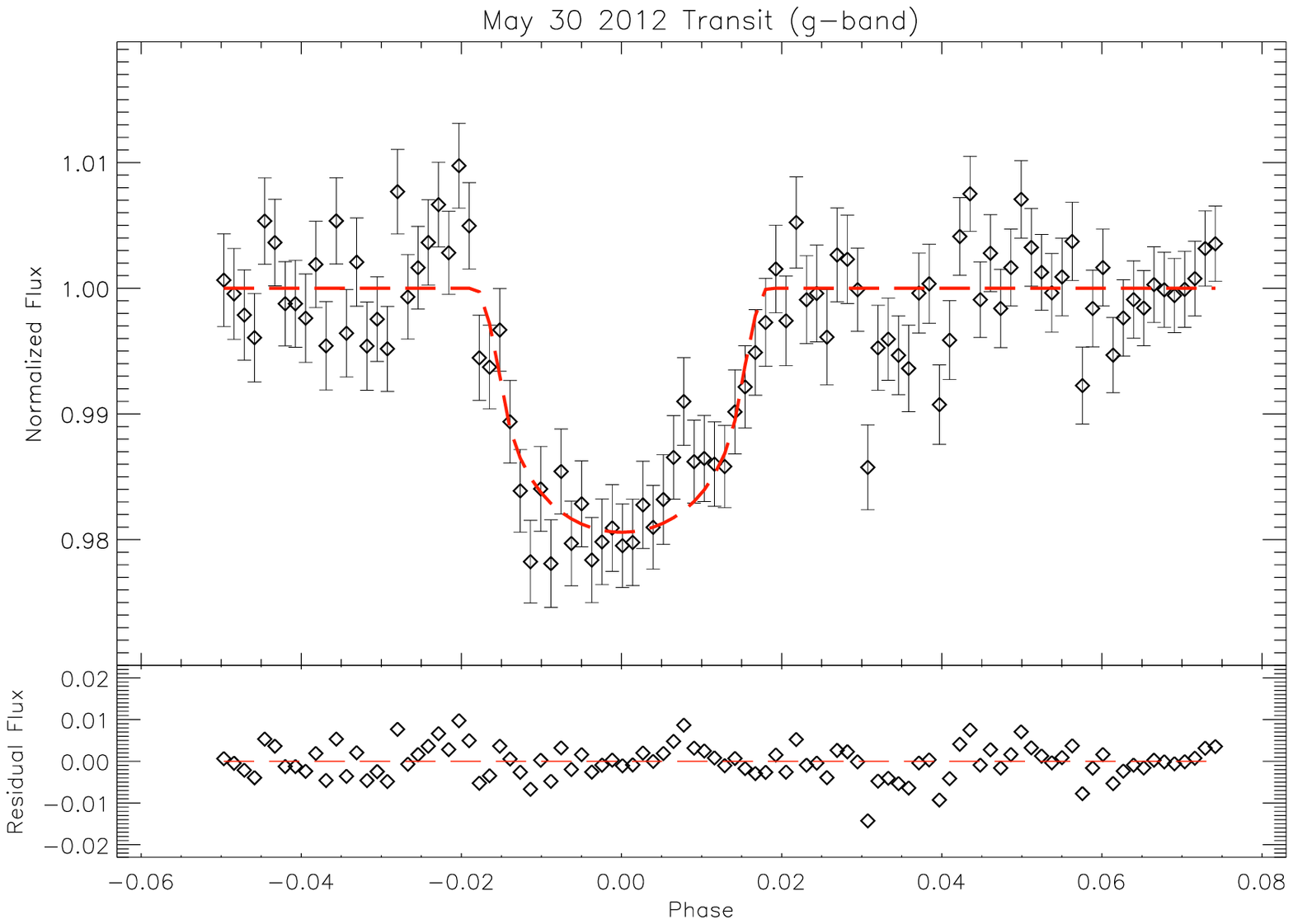}}

  \subfigure{\includegraphics[width=0.5\textwidth]{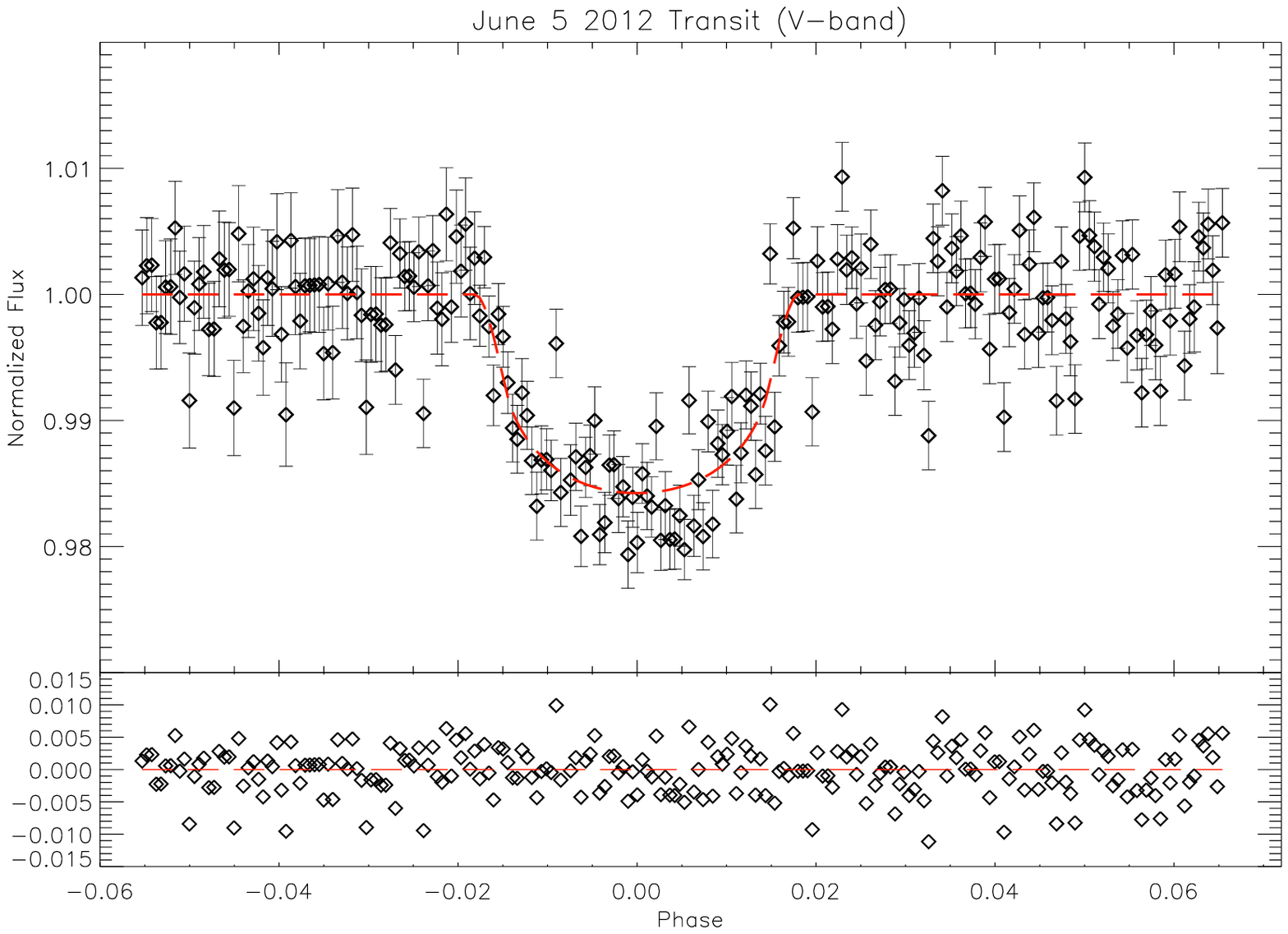}}
\quad
  \subfigure{\includegraphics[width=0.5\textwidth]{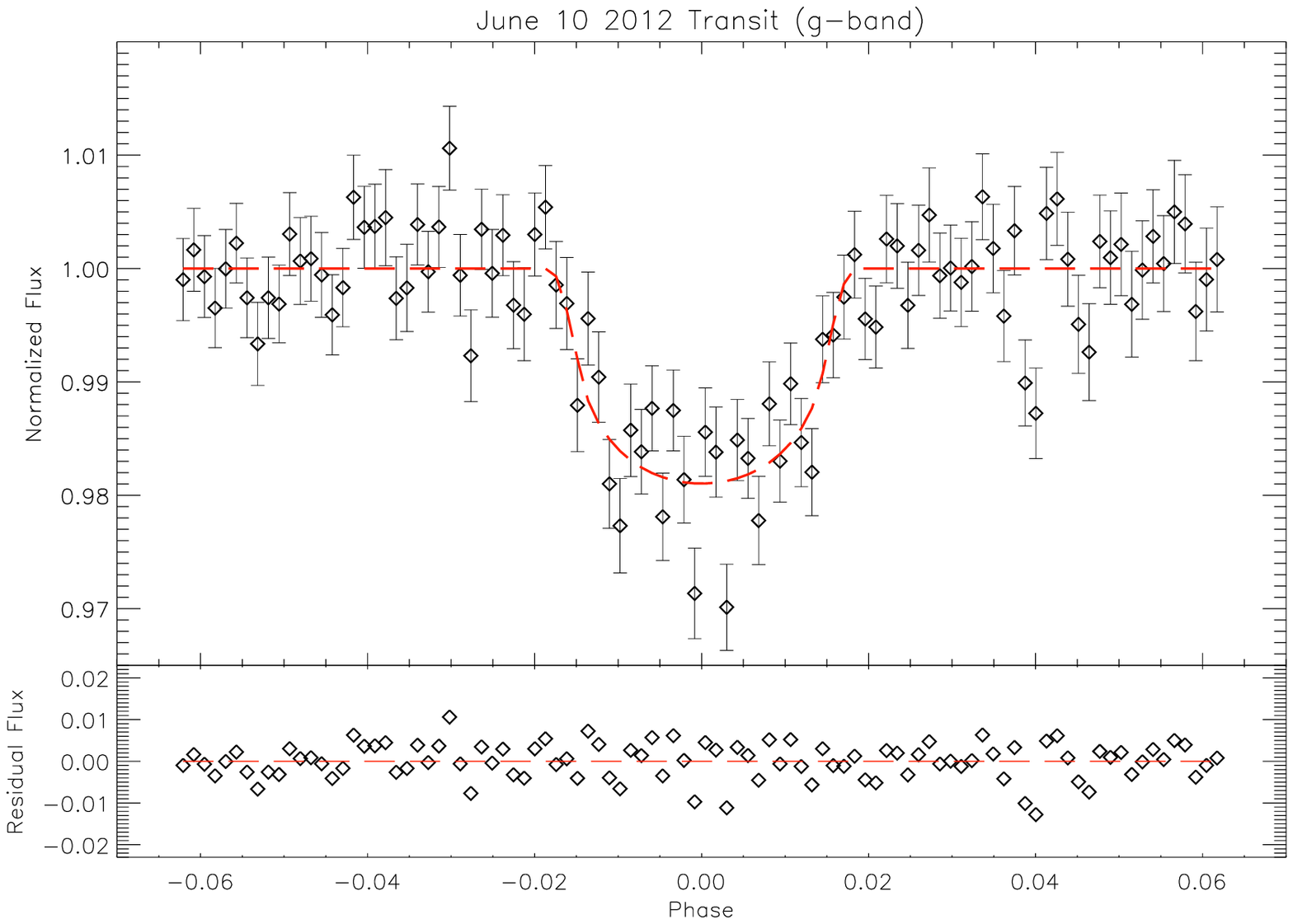}}
\caption{Individual light curves of GJ 1214b for each date of observed
  transit (UTC), shown in chronological order (L-R; top-bottom). The
  data have all been normalised to one, and the linear trend derived
  from the TAP analysis removed. Overplotted with a red dashed line
  are the TAP analysis fits to the data. The residuals from the TAP
  analyses are shown in the lower panels of each plot. The 1$\sigma$ error bars
  plotted on each point are based on the IRAF or SExtractor reduction and were not included in
  the TAP analysis fits.}
\label{gj1214b_lcs2.1}
\end{figure}

\begin{figure}[ht!]
\figurenum{1.2}
   \subfigure{\includegraphics[width=0.5\textwidth]{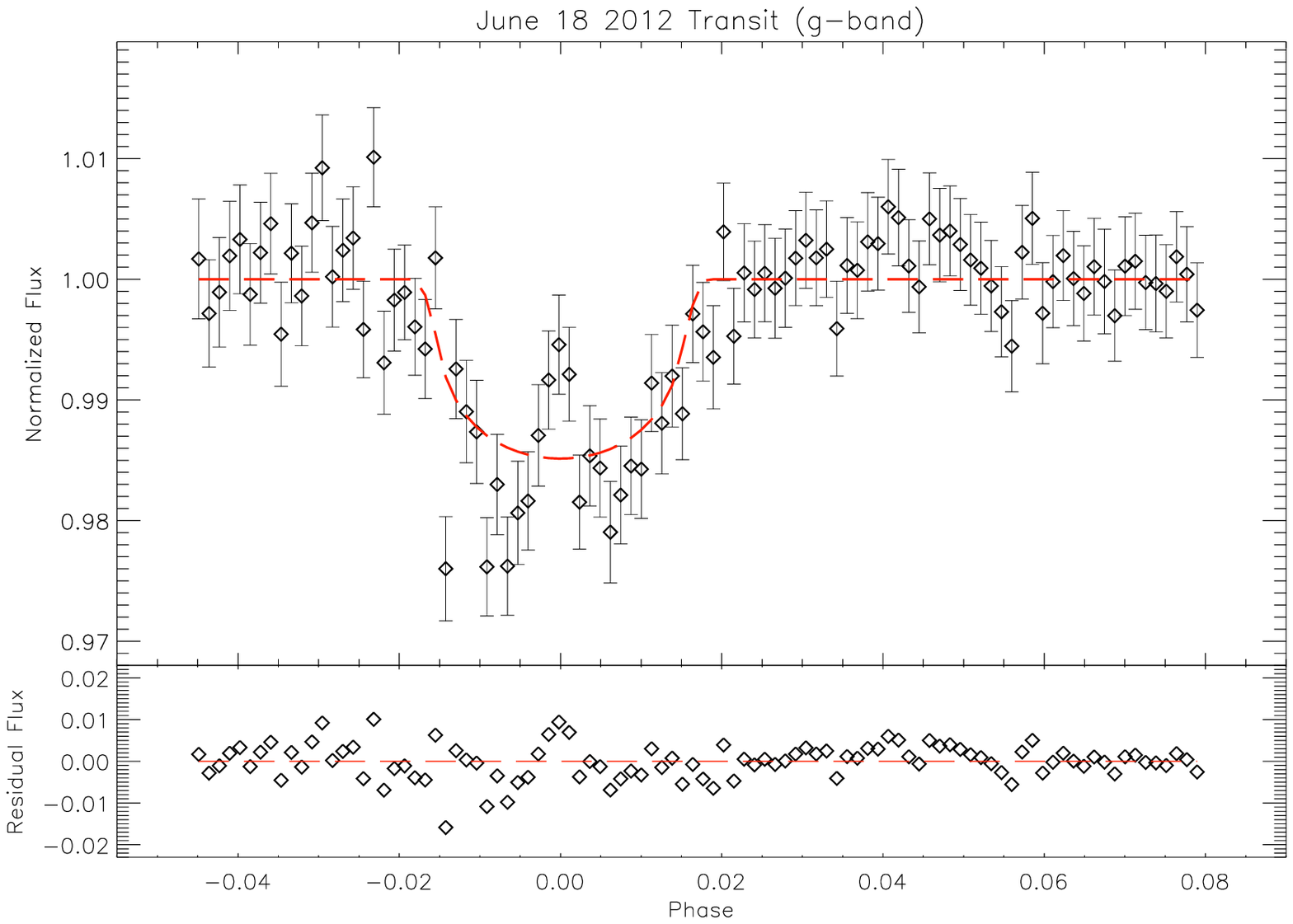}}
\quad
   \subfigure{\includegraphics[width=0.5\textwidth]{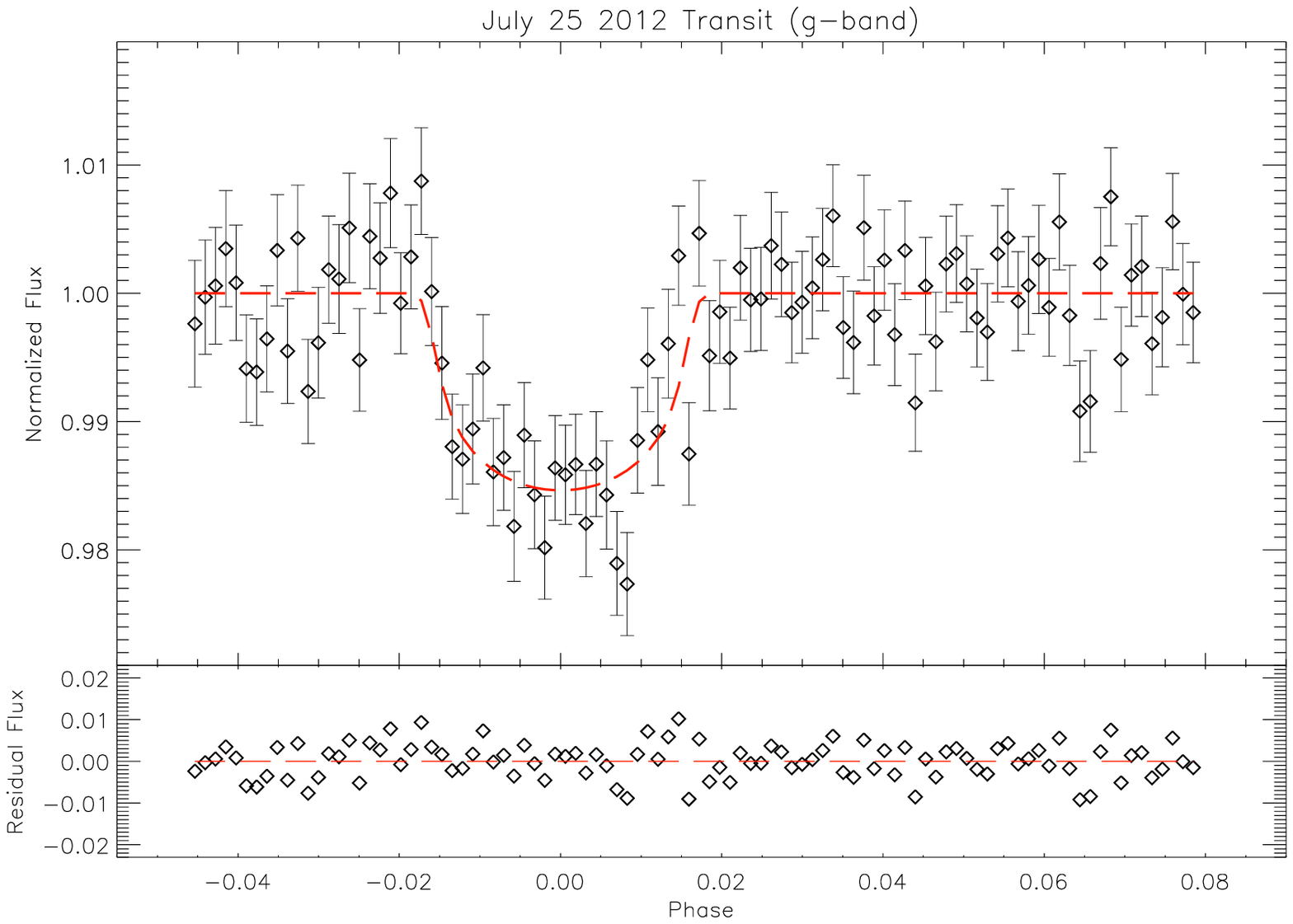}}

   \subfigure{\includegraphics[width=0.5\textwidth]{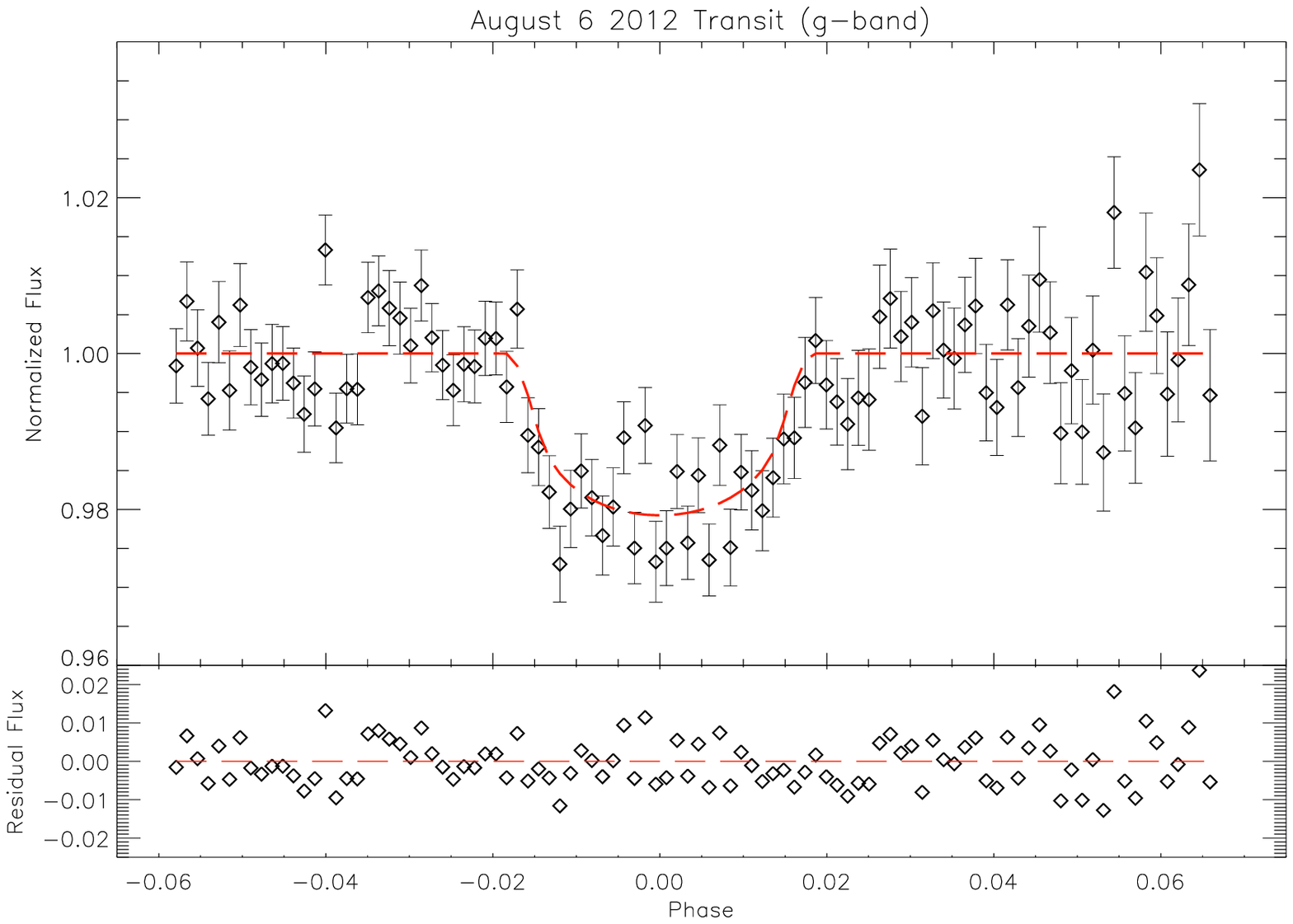}}
\caption{Individual light curves of GJ 1214b for each date of observed
  transit (UTC), shown in chronological order (L-R; top-bottom). The
  data have all been normalised to one, and the linear trend derived
  from the TAP analysis removed. Overplotted with a red dashed line
  are the TAP analysis fits to the data. The residuals from the TAP
  analyses are shown in the lower panels of each plot. The 1$\sigma$ error bars
  plotted on each point are based on the IRAF or SExtractor reduction
  and were not included in the TAP analysis fits.}
\label{gj1214b_lcs2.2}
\end{figure}

To produce the light curves from the Kuiper 1.55 m data, we performed
aperture photometry (using the task PHOT in the
IRAF DAOPHOT package) by measuring the flux from our target star as well
as the flux from several (usually between 5-10) companion stars within
an aperture radius that varied based on the star and the observing
night conditions. For the analysis of each night's observations we
  used a constant sky annulus (with a width of 20 pixels), which was
  chosen to always start at a radius greater (by at least 7 pixels)
  than the target aperture; no stray light from the star was
  included. Considering several different combinations of reference
stars and aperture radii, we picked the combination that produced the
lowest RMS in the out-of-transit data points. To check that our
derived transit depth from the Kuiper 1.55 m data was not
dependent on the chosen aperture radius, we also tested several
different aperture radii and found that the resulting change in
transit depth was not significant based on our derived errors,
which are a factor of $\sim$2 larger. A synthetic light curve was
produced by averaging the light curves from our reference stars, and
the final light curves of GJ 1214b were normalised by dividing by this
synthetic light curve. The light curves for all of our data are shown
in Figures \ref{gj1214b_lcs2.1} and \ref{gj1214b_lcs2.2} with
1$\sigma$ errors on each point converted from magnitude errors
provided by the IRAF reduction. The out-of-transit baseline in all transits achieved a photometric RMS between 2-4 mmag ($\sim$2.5$\times$ the photon noise limit), which is typical for the Mont4k on the 1.55 meter Kuiper telescope for high S/N transit photometry (Turner et al.\,2013; Dittmann et al.\,2009a, 2009b, 2010, 2012; Scuderi et al.\,2010).
  
For the STELLA-I 1.2 m WiFSIP data, we developed a photometry data reduction
pipeline that is based on ESO MIDAS routines to subtract a bias using
the overscan regions and a 2-d bias structure using a masterbias. The
robotic system cycles through all filters of WiFSIP to take twilight
flat-fields resulting in a time difference between science data and
appropriate flat-field observations of less than 3 days. One master
flat-field based on $\sim$20 flat-field exposures was used for flat-field correction. We performed aperture photometry using the
publicly available software SExtractor\footnote{http://www.astromatic.net/software/sextractor} (Bertin \&
Arnouts\,1996), which supplies several options for aperture photometry; we tested the
estimation of fixed aperture magnitudes and automatic aperture
magnitudes in our pipeline. For both options several aperture widths
were tested to minimize the scatter of the out-of-transit data, and we
consistently find an automatic aperture to yield the lowest RMS
value. Again, to check that our derived transit depth from the STELLA-I 1.2 m WiFSIP data was not dependent on the chosen aperture radius, we also tested several different aperture radii and found that the resulting change in transit depth was not significant based on our derived errors. It should be mentioned that this option does not use a constant
aperture shape over the field of view nor the same aperture shape and
width throughout the exposure time series. It computes an elliptical
aperture for every exposure and object by second order moments of its
light distribution (see also Law et al.\,2013; Matute et al.\,2012; Polishook et al.\,2012). Several widths of the `rectangular annulus' used
for local background estimation by SExtractor were tested in order to
minimze the out-of-transit scatter. The same criterium was also used
in the pipeline to search automatically for the best combination of
comparison stars. We always started with the 25 brightest stars in the
field and found 4 to 7 calibration stars to give the optimal
solution. The light curves for all of our data are shown in Figures
\ref{gj1214b_lcs2.1} and \ref{gj1214b_lcs2.2} with
1$\sigma$ errors on each point converted from magnitude errors
provided by the SExtractor reduction. The RMS value of the
out-of-transit STELLA-I data is in most cases $\sim$1.2$\times$ higher than the theoretical limit estimated from the photon noise of object and background and the read-out noise.

\section{Light Curve Analysis}

The light curve depth is a measurement of the effective area of light
from the primary star that is blocked by the occulting planet (($ {
  \frac{R_{p} } {R_{S} } } $)$^{2}$). The effective size of the planet
depends on the opacity of the atmosphere, and thus the atmosphere's
spectral features and composition. To derive the light curve depths,
we used two different publicly available modeling software packages
$-$ the Transit Analysis
Package\footnote{http://ifa.hawaii.edu/users/zgazak/IfA/TAP.html}
(TAP; Gazak et al. 2012) and
JKTEBOP\footnote{http://www.astro.keele.ac.uk/jkt/codes/jktebop.html}
(Southworth et al.\,2004a, 2004b; Southworth 2008) $-$ that simulate
the shape of the light curves, considering the planet's orbit and the limb darkening of the star. TAP utilizes Bayesian probability distributions with Markov Chain
Monte Carlo (MCMC) techniques and a Gibbs sampler to fit transit light
curves using the Mandel \& Agol (2002) model and uses a wavelet
likelihood function to more robustly estimate parameter uncertainties
(Carter \& Winn\,2009). JKTEBOP was originally developed from the EBOP
program written for eclipsing binary star systems (Etzel 1981; Popper
\& Etzel 1981) and uses the Levenberg-Marquadt Monte Carlo (LMMC)
technique to compute errors, although there are additional error
computation options (Southworth et al.\,2004a, 2004b; Southworth 2010; Hoyer et al.\,2011). 

We modeled each transit individually with TAP, after normalizing the
out-of-transit data to one, using five MCMC chains with lengths of
100,000 links each. (We note that TAP does not take into account the 1$\sigma$
individual-point errors as input.) The Gelman-Rubin statistic (Gelman \& Rubin 1992) was used to ensure chain convergence, as outlined in Ford 2006. We also combined the data from the same bands into one simultaneous TAP analysis for each band in order to increase our
sampling and precision; all of our TAP results are listed in Table 3. During the analysis, the inclination (${i}$), scaled semi-major axis ($ { \frac{a } {R_{S} } } $), eccentricity ($e$), argument of periastron ($\omega$), quadratic limb darkening coefficients ($\mu_1$ and $\mu_2$), and the orbital period ($P_{b}$) of the planet were fixed to the values listed in Table 2. The time of mid-transit ($T_c$) and planet-to-star radius ratio ($ { \frac{R_{p} } {R_{S} } } $) were left as free parameters. In addition, white and red noise were left as free parameters, as were the airmass fitting parameters (slope and y-intercept). The linear ($\mu_1$) and quadratic ($\mu_2$) limb darkening coefficients in each respective band were taken from Claret (1998) using approximations of the stellar parameters of GJ 1214 (T$_{eff}$=3000, log $g$=5.0). See Table 2 for the limb darkening coefficients used for each band. 

We also performed a similar analysis of our data with JKTEBOP in order
to check our TAP results against a different transit analysis package. We obtained results consistent with those from TAP, although with slightly smaller errors ($\sim$1.5-3$\times$ smaller). Both JKTEBOP and TAP have been shown to
produce similar results in the study of another transiting exoplanet, WASP-5b (Hoyer, Rojo \& L{\'o}pez-Morales 2012). Hoyer et al.\,(2012) found that, in its default mode, JKTEBOP can underestimate the errors in the fitted parameters because it lacks multi-parameter uncertainty
estimation and does not account for red noise. By including the wavelet
decomposition likelihood function (see Carter \& Winn 2009), TAP
allows parameters that measure photometric scatter (uncorrelated white
noise and time-correlated red noise) to evolve as free parameters in
the transit fitting; the TAP method will recover the traditional
$\chi^{2}$ fitting statistic in the case of no red noise and the white
noise fixed to the characteristic measurement error (Johnson et
al.\,2011). Hoyer et al.\,(2011) also found that if the parameter space does
not have local minima, the LMMC (JKTEBOP)
and MCMC (TAP) algorithms are equivalent, but that LMMC minimization can get trapped in
such minima, and that the LMMC results can be biased toward their
initial input values. We find similar results as Hoyer et al.\,(2012)
in that the errors derived from our TAP analysis are slightly greater than the
errors derived from our JKTEBOP analysis. We choose to use our TAP results throughout the rest of the paper due to their more conservative errors. 

In the June 18 light curve, there appears to be a feature in the
middle of the transit that could affect the TAP analysis and our measurement of $ {
  \frac{R_{p} } {R_{S} } } $. We tested how the model would change by
just excluding these potentially-anomalous data points and performing
our light curve fitting  without them. We do indeed find slightly
larger $ { \frac{R_{p} } {R_{S} } } $ values for the transit of June 18, which in turn slightly increases our combined-night $ { \frac{R_{p} } {R_{S} } } $ value in $g'$-band (by $\sim$0.0014). However, within errors, these values are consistent with the values we derive using all of the data points; to avoid any bias due to attempts to fit out these points, we report here the values derived using all of the data.

Our TAP analysis results are summarized in Table \ref{tab3}, and a comparison of our results and $ { \frac{R_{p} } {R_{S} } } $ values from the literature is shown in Figures \ref{all_2ptresults_withothers} and \ref{all_2ptresults_withothers_zoomout}.

\section{Discussion \& Conclusions} 
\begin{center}
\begin{figure}[ht!]
\figurenum{2}
\centering
\includegraphics[width=6.8in]{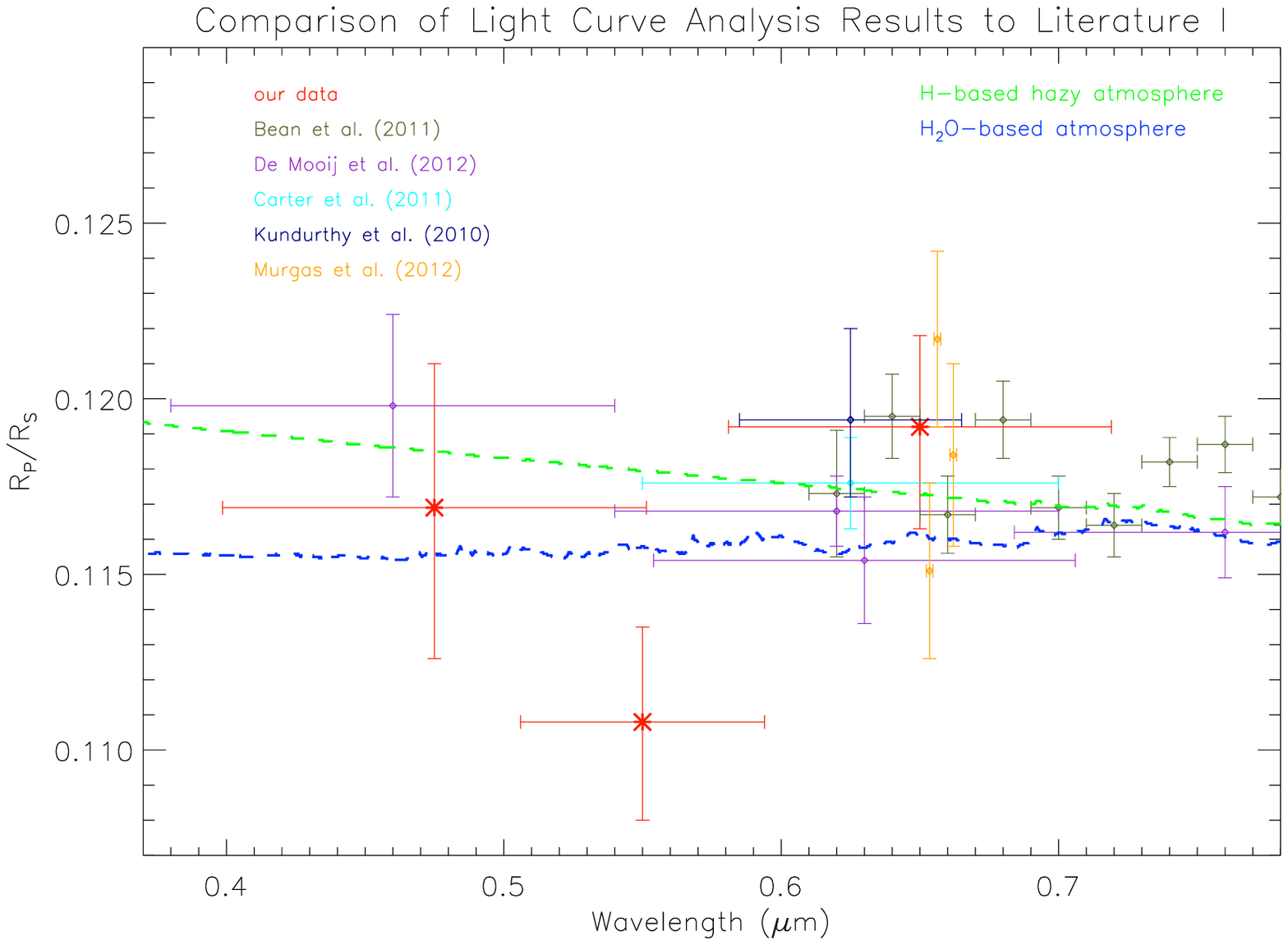}
\caption{The results of our combined-night analyses (the last three
  rows in Table \ref{tab3}), as compared to other published transit measurements of GJ 1214b. Our results are bolded in red; the band-pass error bars on our measurements represent the FWHM of each filter. We overplot two examples of end-member models that are consistent with different selections of the existing data: a hydrogen-based and hazy atmosphere with a solar abundance of water (green) and a water-based atmosphere (blue). The former displays a greater modulation in the spectral features as a result of the higher scale height of the H$_2$-based atmosphere.}
\label{all_2ptresults_withothers}
\end{figure}
\end{center}

Our derived $ { \frac{R_{p} } {R_{S} } } $ values match those in the
literature (see Figure \ref{all_2ptresults_withothers}). Our analysis
adopts the same values for the period, inclination ($i$), $ { \frac{a
  } {R_{S} } } $, eccentricity, and omega (see Table 2) used by Bean et al.\,(2010) and Bean et al.\,(2011), making our results directly
comparable to theirs. These values were also used by de Mooij et
al.\,(2012), except for the period, for which de Mooij et al.\,(2012)
used 1.5804048346 days rather than Bean et al.\,(2011)'s 1.58040481
days. Bean et al.\,(2011) allowed the limb darkening coefficients to
be free parameters in their fitting analysis, using as priors the
theoretical values that they computed based on PHOENIX models of GJ
1214A with stellar parameters T$_{eff}$ $=$3026 K, [M/H] $=$0.0, and
log $g$ $=$5.0. Neither the theoretical priors nor the resulting
fitted values for the limb-darkening coefficients are discussed in
Bean et al.\,(2011), so we cannot compare our limb-darkening
coefficients directly. We did use the limb darkening coefficient
values from various Claret sources (see Table 2)
corresponding to stellar parameters very similar to Bean et
al.\,(2011): T$_{eff}$ $=$3000 K, [M/H] $=$0.0, and log $g$ $=$5.0. De
Mooij et al.\,(2012) used a four-parameter limb-darkening law, so our
coefficients are also not directly comparable, although de
Mooij et al.\,(2012) do use the same stellar parameter values (T$_{eff}$ $=$3026 K, [M/H] $=$0.0, and log $g$ $=$5.0) and Claret (2000; 2004) as sources for their non-linear limb-darkening coefficients.

GJ 1214A is known to have star-spot-induced variability (Charbonneau
et al.\,2009; Berta et al.\,2011), and stellar activity can have an
observable effect on the transmission spectrum of a transiting planet
from star-spots that are occulted or not occulted by the planet (Pont
et al.\,2008; Agol et al.\,2010; Sing et al.\,2011). If a planet
passes in front of a star-spot, fully or partially masking it on the
stellar surface, the observed flux will increase in proportion to the
dimming effect of the star-spot on the total flux of the star, causing
one to underestimate the true size of the planet, and decreasing $ { \frac{R_{p}
  } {R_{S} } } $. If the star spot is not occulted by the planet, the
transit depth will appear greater, since the planet will pass over a
region that is on average brighter than the entire star; this will
reduce the effective stellar radius and increase $ { \frac{R_{p} }
  {R_{S} } } $ (these effects are parameterized in Sing et
al.\,2011). De Mooij et al.\,(2012) found from their out-of-transit
monitoring observations of GJ 1214A that the corrections in their $ {
  \frac{R_{p} } {R_{S} } } $ observations in $r$-band and $K_s$-band
due to the possibility of occulted star-spots were 0.0011 and 0.0003,
respectively. These authors also calculated the influence of different
base levels of unocculted spots on the transmission spectrum
observations of GJ 1214b, and found that for a spot-covering fraction
of 10\%, the change in their $ { \frac{R_{p} } {R_{S} } } $ values was
-0.0007 in $g$-band and $\sim$-0.00055 in $r$-band (the authors
shifted their values such that the $i$-band radius ratio remained the
unaltered baseline value, to compare to non-corrected results more easily). Due to the errors in our derived $ { \frac{R_{p} } {R_{S} } } $ values, the level of star-spot-induced variations calculated by de Mooij et al.\,(2012) is not distinguishable with our data. 

\begin{center}
\begin{figure}[ht!]
\figurenum{4}
\centering
\includegraphics[width=6.8in]{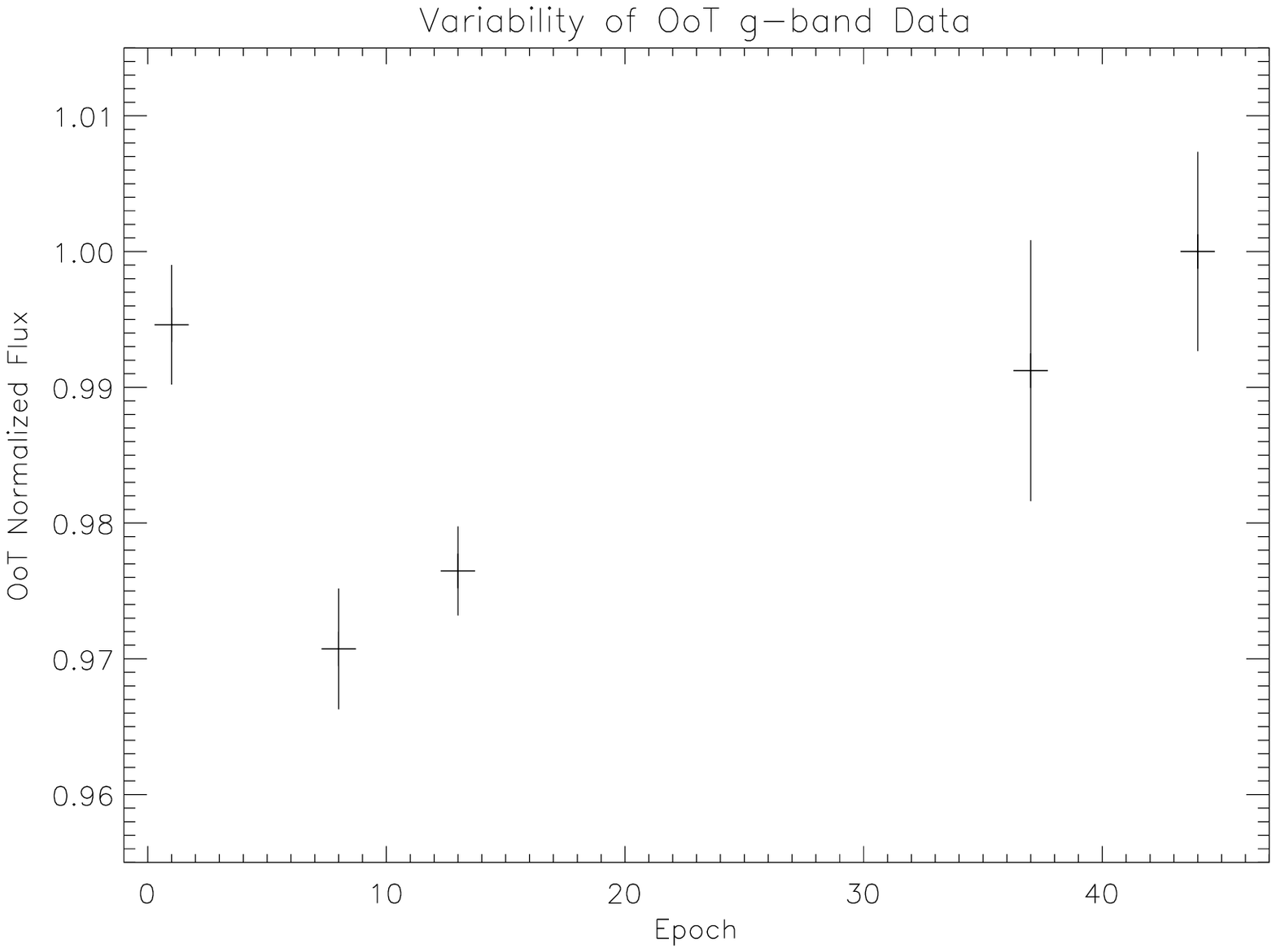}
\caption{The variability in the out-of-transit (OoT) $g'$-band data from our observations. The x-axis represents the normalised epoch at which the data were observed, with the first $g'$-band epoch set to one. The out-of-transit data from each night were calibrated using the same set of comparison stars; then we found the mean out-of-transit flux level of each night and normalised all the data by dividing through the highest mean value, corresponding to the time at which the stellar surface was least spotted. The greatest difference is between epochs 8 (June 10) and 44 (August 6), $\sim$3\%. The error bars represent the standard deviation of the out-of-transit flux for each night}
\label{gband_ootvar}
\end{figure}
\end{center}

We also performed our own check calculations for possible star-spot
corrections, using our $g'$-band out-of-transit data, i.e., data taken
on the same nights as (and acting as the baseline for) the transit
data. This check allows us to directly probe the host star variability
in $g'$-band, our `bluest' band and thus the one most affected by
spots. We took the out-of-transit data from each night, calibrated
using the same comparison stars, found the mean relative
out-of-transit flux, and normalised it to the brightest epoch (August
6; see Figure \ref{gband_ootvar}), which we assume to be the epoch
with the lowest spot coverage. Between the dimmest (June 10) and
brightest (August 6) epochs, there was a change of $\sim$3\% in the
flux of the star, translating to a transit depth that is deeper by
1/0.97, or $\sim$1.03, due to star-spots. To ensure that the observed
change in flux of GJ 1214A was not due to systematic error, we
performed the same analysis on three of the closest (in angular
  separation) reference stars and found that their flux varied by $<$1\% over the time period of our $g'$-band observations. Thus we assume that the $\sim$3\% variability of GJ 1214A's flux is real, and that this change in host star flux is due to dark spots; we do not consider bright plage or faculae regions. Note that a dark region that is unocculted will make the transit appear deeper than it really is. Applying this star-spot correction to the June 10 transit results in a transit depth correction of (0.1196)$^{2}\times$0.03 $=$ 0.00043, or an $ { \frac{R_{p} } {R_{S} } } $ correction of 0.0018 (using our TAP-analysis values for June 10; see Table \ref{tab3}). According to our data, this is the \textit{greatest} magnitude of correction that could affect our $ { \frac{R_{p} } {R_{S} } } $ values, and it is markedly less than our TAP-based $ { \frac{R_{p} } {R_{S} } } $ errors on June 10 of $^{+0.0064}_{-0.0068}$. Since June 10 was the dimmest epoch, on the other $g'$-band nights the potential star-spot corrections are even smaller, and for $V$- and $R$-bands we can assume a lower flux variation due to the lower flux contrast between spots and the surrounding stellar surface at redder wavelengths. So, while the variability of GJ 1214A should be taken into account when evaluating transit observations taken over multiple epochs, we confirm that the resulting difference in $ { \frac{R_{p} } {R_{S} } } $ that could be induced by star-spots is well within our error bars and thus not distinguishable with our observations.

\begin{center}
\begin{figure}[ht!]
\figurenum{3}
\centering
\includegraphics[width=6.8in]{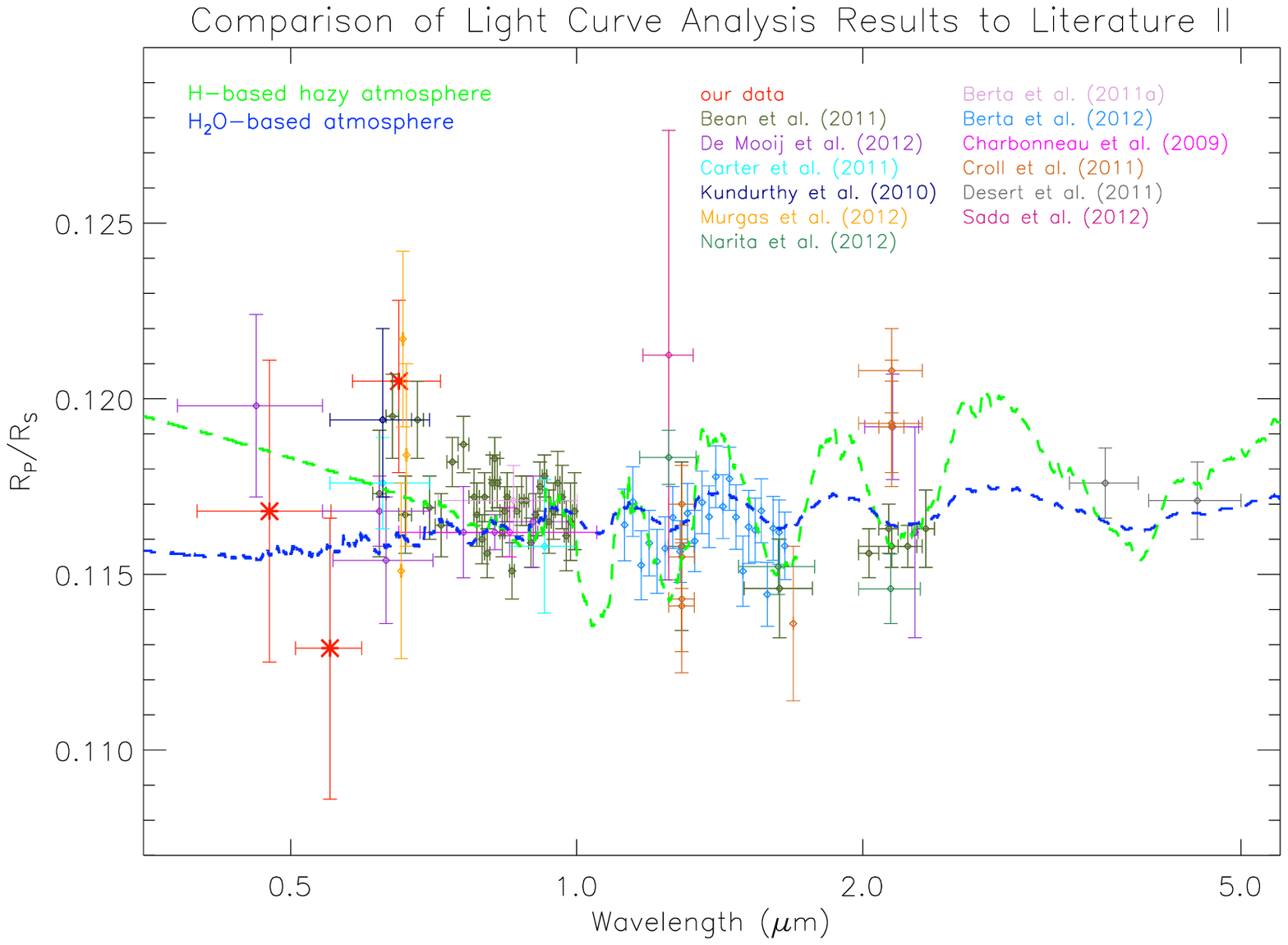}
\caption{The same as Figure \ref{all_2ptresults_withothers}, but now including data covering a greater range in wavelength. See caption of Figure 3 for details.}
\label{all_2ptresults_withothers_zoomout}
\end{figure}
\end{center}

We find agreement within errors between our data, based on five nights
of observations, and the large $g'$-band planet radius found by de
Mooij et al.\,(2012), which was based on only one night of
observing. However, our combined $g'$-band observations (last line in Table 3)
show that the $g'$-band planet radius could actually be smaller ($\sim$0.7$\sigma$ shallower) than that found by de Mooij et
al.\,(2012). Taken with the low $V$-band $ { \frac{R_{p} } {R_{S} } }
$ value that we find (where $V$-band spans 473-686 $\mu$m; FWHM 88
$\mu$m), our results suggest that the planet-to-star radius ratio does not
increase significantly at shorter wavelengths; within our TAP-analysis-derived errors, the spectrum of GJ 1214b is consistent with zero slope (flat) from $\sim$400-800 nm (see Figures \ref{all_2ptresults_withothers} and \ref{all_2ptresults_withothers_zoomout}).

Current transmission observations of GJ 1214b are somewhat
contradictory at optical and $K$-band wavelengths, which complicates
studies of its composition. While most high signal-to-noise
observations indicate a featureless, flat spectrum across the optical
and near-infrared (Berta et al.\,2012; Bean et al.\,2010; Bean et al.\,2011;
D{\'e}sert et al.\,2011; Narita et al.\,2012), the measurements of Croll et al.\,(2011)
(in $K_s$-band), de Mooij et al.\,(2012) ($g$-band and $K_s$-band),
and Murgas et al.\,(2012) (around $R$-band) indicate potential variation in transit depth with wavelength. Yet there are a few constraints that persist, considering the two end-member models, one that is hydrogen-based and another that is water-(or heavy gas) based, that have been proposed to explain the structure of GJ 1214b's atmosphere. The observed spectral features of GJ 1214b are sufficiently muted such that, if it did have an H$_2$ rich atmosphere, the prominent spectral features of water would need to be reduced by adding large sized ($>$1 micron) particulates (Bean et al.\,2010; Croll et al.\,2011; Berta et al.\,2012; Howe \& Burrows 2012) and/or reducing the water abundance to one lower than that expected in a solar elemental abundance atmosphere (de Mooij et al.\,2012; Howe \& Burrows 2012). Alternatively the variations on a water-rich atmosphere proposed for GJ 1214b have muted features as a result of the atmosphere's larger mean molecular mass and thus smaller scale height. These models match most of the spectra except the high absorption measured in $K$-band (Croll et al.\,2011; de Mooij et al.\,2012) and arguably in $g$-band (de Mooij et al.\,2012).

Here we investigate one of the largest differences between the
H$_{2}$O-based and H$_2$-based atmospheres $-$ their spectral
signatures at optical wavelengths. To illustrate the disparity in the
spectra predicted for these two end-member atmospheric structures, we
calculate spectra of an H$_2$ and an H$_2$O atmosphere, consistent with prior studies. Transit
depths $ ({ \frac{R_{p} } {R_{S} } })^2$ were calculated with a
numerical model that sums the contributions of the primary star's
transmission through the limb of the extrasolar planet. The absorption
of light is derived along tanget lines at pressures that extend from
10$^{-7}$ bars to 10 bars. Since there is no evidence so far of the
presence of methane or ammonia, as would be expected in a thermochemical
equilibrium atmosphere at the temperatures in GJ 1214b's atmosphere
(Miller-Ricci Kempton et al.\,2012), we include the spectroscopic
absorption due to water only, which is calculated using the absorption
coefficients of Freedman, Marley \& Lodders\,(2008), and assuming a constant
mixing ratio, as expected for the pressure levels we are probing
(below the 10$^{-5}$ bar level) (Miller-Ricci Kepmton et
al.\,2012). The H$_2$-based model shown in Figures
\ref{all_2ptresults_withothers} and
\ref{all_2ptresults_withothers_zoomout} assumes a water abundance of
3.5$\times10^{-5}$ and a cloud of brightly scattering particles (with
real and imaginary indices of refraction of 1.65 and 10$^{-4}$) below
1 mbar, which represents one solution that mutes the water
features. This particular model, one of many degenerate solutions, is
compared to that of a non-cloudy water atmosphere to illustrate the
different slopes between 0.3-0.9 $\mu$m that result primarily from the
different atmospheric scale heights. The H$_2$-based atmosphere has a
spectrum that demostrates the increase in opacity due to Rayleigh
scattering, which is suggested by the observations of de Mooij et
al.\,(2012). The H$_{2}$O-based atmosphere is excluded by the
observations of de Mooij et al.\,(2012), because the small scale
height of the model depresses the 0.46 $\mu$m radius below that measured.

Our measurements are consistent with prior studies; we measure a
$g'$-band radius that agrees with de Mooij et al.\,(2012), but allows
for a greater number of solutions that include an H$_{2}$O-based
atmosphere. We have recorded the first $V$-band observations of GJ
1214b (centered at 0.55 $\mu$m). These data point to a lower
absorption more consistent with an H$_{2}$O-rich atmosphere or a
mixture of H$_2$ and water; that is, an intermediate atmospheric
structure. Such an atmosphere might be expected because any
H$_{2}$O-rich atmosphere would necessarily produce hydrogen through
photochemistry. We measure an $R$-band radius that is also consistent with either a H$_2$ or an intermediate water and H$_2$-based atmosphere. Taken together, our observations can be best interpreted with an atmosphere that is partly H$_2$ and partly water based. However, additional observations are needed from ground-based and space-based platforms to establish the optical continuum of GJ 1214b. 
   
\section{Acknowledgments}
\acknowledgments 
Data presented in this paper were obtained with the
STELLA robotic telescope in Tenerife, an AIP facility jointly operated
by AIP and IAC. We are grateful to the staff of the Kuiper 1.55 m
telescope for their patience and the University of Arizona TAC for
awarding time for these observations. The authors thank Amy Robertson,
Timothy Carleton, and Kevin Hardegree-Ullman for their help observing
at the Kuiper telescope; Thomas Granzer for his
STELLA support and Klaus G. Strassmeier for his advice; J. Southworth
and J. Eastman for productive conversations about their exoplanet
transit fitting programs; and the anonymous referee for his/her
helpful edits and comments. C. Griffith, J. Turner, and J. Teske were
partially supported by NASA Planetary Astmopheres Grant No. NNX11AD92G. 

NOTE: In the process of resubmission of this work, another paper
concerning GJ 1214b was submitted (Fraine et al.\,2013). The conclusions of the present work are unaffected by the new submission. 



\begin{landscape}
\begin{deluxetable}{cccccccc}
\tablenum{1}
\tablecolumns{8}
\tablewidth{0pc} 
\tabletypesize{\footnotesize}
\tablecaption{Summary of Observations \label{tab1}}
\tablewidth{0pt} 
\tablehead{
\colhead{Observing Night}            & 
\colhead{Filter} &
\colhead{Telescope} &
\colhead{Start-Stop} &
\colhead{Int. Time}  &
\colhead{In-Transit/Total Frames} &
\colhead{Seeing} & 
\colhead{Out of Transit RMS} \\ 
\colhead{(UTC)}            & 
\colhead{} &
\colhead{} &
\colhead{(UTC)} &
\colhead{(s)}  &
\colhead{} &
\colhead{(arsec)} & 
\colhead{(mmag)}
}
\startdata 
March 28-29 2012 & Harris R & Kuiper $1.55$ m & 08:04-11:23 &  50 & 54/204 & 1.5-2.6 & 2.05\\  
April 8-9 2012 &   Harris R   &  Kuiper $1.55$ m & 10:25-11:04&  50 & 51/156 & 1.2-2.0 & 4.21 \\
May 5-6 2012   &   Harris V & Kuiper $1.55$ m & 06:47-09:41 &  100 & 23/73 &1.8-2.6 & 2.97  \\
May 29-30 2012  & Sloan $g'$ & STELLA $.2$ m  & 23:52-02:51   &  90 & 29/98 & 1.09-1.24  & 4.11 \\ 
June 4-5 2012 &   Harris V & Kuiper $1.55$ m  & 07:30-10:20  &  30 & 69/235 & 1.0-1.9&  3.97  \\
June 9-10 2012  & Sloan $g'$ & STELLA $1.2$ m & 01:03-04:02    &  90 &29/98 & 1.09-1.48   & 4.14 \\ 
June 17-18 2012 & Sloan $g'$ & STELLA $1.2$ m & 23:05-02:04    &  90 &29/98 & 1.09-1.17   & 3.22 \\
July 25-26 2012 & Sloan $g'$ & STELLA $1.2$ m  & 21:27-00:27   &  90 &28/98 & 1.09-1.32    & 4.02 \\
August 5-6 2012 & Sloan $g'$ & STELLA $1.2$ m  & 22:40-01:40   &  90  & 29/98 & 1.14-2.37  & 5.50 (for airmass$<$2)\\
\enddata
\tablecomments{Column 8 gives the Out-of-Transit root-mean-squared (RMS) relative flux.
}
\end{deluxetable} 
\end{landscape}

\begin{table} [ht]
\centering
\begin{tabular}{ccc}
\tablenum{2} 
 & Table 2. Fixed Model Values for TAP &  \\ 
\hline
Period &  1.58040481 & Bean et al.\,2011        \\
Inclination& 88.94  & Bean et al.\,2011         \\
a/R$_{S}$&  14.9749        & Bean et al.\,2011  \\
Eccentricity&  0.0   &     \\
Omega&        0.0 & \\
Harris V limb darkening coefficients & 0.6406, 0.2955 & Claret 1998 \\
Harris R limb darkening coefficients & 0.5392, 0.3485  & Claret 1998 \\
Sloan $g'$ limb darkening coefficients & 0.6528, 0.2978 & Claret 2004 \\
\hline
\end{tabular}
\label{tab2}
\end{table}

\begin{deluxetable}{c c c c c c c }
\tablenum{3}
\tablecolumns{7}
\tabletypesize{\footnotesize}
\tablecaption{TAP Model Fitting Results \label{tab3}}
\tablewidth{0pt} 
\tablehead{
\colhead{Transit Date} &
\colhead{Filter} & 
\colhead{Midtransit Time}            &
\colhead{R$_p$/R$_S$} &
\colhead{Airmass slope} &
\colhead{Airmass y-intercept} &
\colhead{Red Noise} \\
\colhead{(UTC)} &
\colhead{} & 
\colhead{(BJD)}            &
\colhead{} &
\colhead{} &
\colhead{} &
\colhead{}
}
\startdata
March 29 &  Harris R   & 2456015.91453 $^{+0.00037}_{-0.00035}$ & 0.1203 $^{+0.0027}_{-0.0030}$ & -0.0098 $^{+0.0094}_{-0.010}$ & 1.0005 $^{+0.00089}_{-0.00084}$ & 0.0073 $^{+0.0032}_{-0.0026}$ \\ 
April 9 & Harris R  & 2456026.97465 $^{+0.00050}_{-0.00051}$ & 0.1192 $^{+0.0037}_{-0.0040}$ & -0.0340 $^{+0.018}_{-0.018}$ & 1.0002 $^{+0.0012}_{-0.0011}$ & 0.0072 $^{+0.0055}_{-0.0047}$\\
May 6 &  Harris V  &  2456053.84203 $^{+0.00063}_{-0.00064}$ & 0.1108 $^{+0.0069}_{-0.0088}$ & -0.0240 $^{+0.021}_{-0.023}$ & 1.0011 $^{+0.0018}_{-0.0016}$ & 0.0098 $^{+0.0054}_{-0.0046}$\\
May 30 & Sloan $g'$ & 2456077.54970 $^{+0.0011}_{-0.0012}$   & 0.1210 $^{+0.0096}_{-0.011}$  & 0.0070 $^{+0.031}_{-0.030}$  & 0.9995 $^{+0.0025}_{-0.0026}$   & 0.0171 $^{+0.0067}_{-0.0058}$  \\
June 5 &  Harris V &  2456083.87044 $^{+0.00058}_{-0.00058}$ & 0.1093 $^{+0.0049}_{-0.0050}$ & -0.0660 $^{+0.015}_{-0.015}$ & 1.0015 $^{+0.0012}_{-0.0012}$ & 0.0083 $^{+0.0061}_{-0.0052}$\\
June 10 & Sloan $g'$ &  2456088.51112 $^{+0.00093}_{-0.00083}$ & 0.1197 $^{+0.0068}_{-0.0070}$  & 0.0040 $^{+0.025}_{-0.023}$   & 0.9995 $^{+0.0018}_{-0.0019}$    & 0.0084 $^{+0.0077}_{-0.0056}$  \\ 
June 18 & Sloan $g'$ & 2456096.50120 $^{+0.0014}_{-0.0015}$ &  0.1058 $^{+0.0096}_{-0.012}$ & 0.0310 $^{+0.031}_{-0.031}$   & 0.9966 $^{+0.0026}_{-0.0027}$    & 0.0166 $^{+0.0073}_{-0.0071}$  \\ 
July 25 & Sloan $g'$ & 2456134.44320 $^{+0.0010}_{-0.0011}$ & 0.1077 $^{+0.0078}_{-0.0082}$  &  0.0000 $^{+0.025}_{-0.024}$  &   0.9995 $^{+0.0020}_{-0.0020}$  &  0.0090 $^{+0.0085}_{-0.0059}$ \\ 
August 6 & Sloan $g'$ &  2456145.50590 $^{+0.0012}_{-0.0014}$ & 0.1250 $^{+0.012}_{-0.018}$  &  0.1110 $^{+0.043}_{-0.047}$  & 0.9932 $^{+0.0038}_{-0.0035}$    &  0.022 $^{+0.014}_{-0.012}$ \\ 
\hline
2 nights       & Harris R &          ---                          & 0.1192 $^{+0.0026}_{-0.0029}$ & 0.0000 $^{+0.011}_{-0.011}$ & 1.0000 $^{+0.00093}_{-0.00091}$ &  0.0104 $^{+0.0034}_{-0.0032}$ \\
2 nights       & Harris V &            ---                        & 0.1108 $^{+0.0027}_{-0.0028}$ & 0.0019 $^{+0.0089}_{-0.0089}$ & 0.9998 $^{+0.00068}_{-0.00067}$ & 0.0043 $^{+0.0042}_{-0.0030}$ \\
5 nights       & Sloan $g'$ &           ---                       & 0.1169 $^{+0.0041}_{-0.0043}$ &  0.030 $^{+0.013}_{-0.012}$  & 0.9974 $^{+0.0011}_{-0.0012}$  &  0.0092 $^{+0.0072}_{-0.0060}$  \\ 
\enddata
\end{deluxetable}


\begin{thebibliography}{dummy}

\bibitem{}Agol E., Cowan N.~B., Knutson H.~A., Deming D., Steffen J.~H., Henry G.~W., Charbonneau D., 2010, ApJ, 721, 1861

\bibitem{}Bean J.~L. et al., 2011, ApJ, 743, 92 

\bibitem{}Bean J.~L., Miller-Ricci Kempton E., Homeier D., 2010, \nat, 468, 669 

\bibitem{}Berta Z.~K. et al., 2012, \apj, 747, 35

\bibitem{}Berta Z.~K. et al., 2011, \apj, 736, 12 

\bibitem{}Bertin E., Arnouts S., 1996, \aaps, 117, 393 

\bibitem{}Borucki W.~J. et al., 2012, \apj, 745, 120 

\bibitem{}Borucki W.~J., for the Kepler Team, 2010, arXiv:1006.2799 

\bibitem{}Carter J.~A., Winn J.~N., 2009, \apj, 704, 51

\bibitem{}Charbonneau D. et al., 2009, \nat, 462, 891

\bibitem{}Claret A., Bloemen S., 2011, VizieR Online Data Catalog, 352, 99075 Claret \& Bloemen (2011)

\bibitem{}Croll B., Albert L., Jayawardhana R., Miller-Ricci Kempton E., Fortney J.~J., Murray N., Neilson, H., 2011, \apj, 736, 78

\bibitem{}Crossfield I.~J.~M., Barman T., Hansen B.~M.~S., 2011, \apj, 736, 132 

\bibitem{}De Mooij E.~J.~W. et al., 2012, \aap, 538, A46 

\bibitem{}D{\'e}sert J.-M. et al., 2011, \apjl, 731, L40 

\bibitem{}Dittmann J.~A., Close L.~M., Scuderi L.~J., Turner J., Stephenson P.~C., 2012, \na, 17, 438 

\bibitem{}Dittmann J.~A., Close L.~M., Scuderi L.~J., Morris M.~D., 2010, \apj, 717, 235

\bibitem{}Dittmann J.~A., Close L.~M., Green E.~M., Scuder, L.~J., Males J.~R., 2009a, \apjl, 699, L48 

\bibitem{}Dittmann J.~A., Close L.~M., Green E.~M., Fenwick M., 2009b, \apj, 701, 756 

\bibitem{}Etzel P.~B., 1981, Photometric and Spectroscopic Binary Systems, 111 

\bibitem{}Ford E.~B., 2006, \apj, 642, 505 

\bibitem{}Fraine J.~D., Deming D., Gillon M. et al., 2013, arXiv:1301.6763 

\bibitem{}Freedman R.~S., Marley M.~S., Lodders K., 2008, ApJS, 174, 504

\bibitem{}Gazak J.~Z., Johnson J.~A., Tonry J., Dragomir D., Eastman J., Mann A.~W., Agol E., 2012, Advances in Astronomy, 2012

\bibitem{}Gelman A., Rubin D.~B., 1992, Stat. Sci., 7, 457

\bibitem{}Harps{\o}e K.~B.~W. et al., 2013, \aap, 549, A10

\bibitem{}Howard A.~W. et al., 2010, Science, 330, 653 

\bibitem{}Hoyer S., Rojo P., L{\'o}pez-Morales M., 2012, \apj, 748, 22 

\bibitem{}Hoyer S., Rojo P., L{\'o}pez-Morales M., D{\'i}az R.~F., Chambers J., Minniti D., 2011, \apj, 733, 53 

\bibitem{}Kundurthy P., Agol E., Becker A.~C., Barnes R., Williams B., Mukadam A., 2011, \apj, 731, 123 

\bibitem{}Johnson J.~A. et al., 2011, \apj, 730, 79

\bibitem{}Law N.~M. et al., 2013, \apj, 145, 58

\bibitem{}Mandel K., Agol E., 2002, \apjl, 580, L171

\bibitem{}Matute I. et al., 2012, \aap, 542, A20 

\bibitem{}Miller-Ricci Kempton E., Zahnle K., Fortney J.~J., 2012, \apj, 745, 3

\bibitem{}Miller-Ricci E., Fortney J.~J., 2010, \apjl, 716, L74 

\bibitem{}Murgas F., Pall{\'e} E., Cabrera-Lavers A., Col{\'o}n K.~D., Mar{\'i}n E.~L., Parviainen H., 2012, \aap, 544, A41 

\bibitem{}Muirhead, P.~S. et al., 2012, \apj, 747, 144 

\bibitem{}Narita N., Nagayama T., Suenaga T. et al., 2012, arXiv:1210.3169 

\bibitem{}Polishook D. et al., 2012, \mnras, 241, 2094

\bibitem{}Pont F., Knutson H., Gilliland R.~L., Moutou C., Charbonneau D., 2008, MNRAS, 385, 109

\bibitem{}Popper D.~M., Etzel P.~B., 1981, \aj, 86, 102

\bibitem{}Rogers L.~A., Seager S., 2010, ApJ, 716, 1208

\bibitem{}Sada P.~V. et al., 2012, \pasp, 124, 212 

\bibitem{}Scuderi L.~J., Dittmann J.~A., Males J.~R., Green E.~M., Close L.~M., 2010, \apj, 714, 462 

\bibitem{}Sing D.~K. et al., 2011, MNRAS, 416, 1443

\bibitem{}Southworth J., 2008, \mnras, 386, 1644

\bibitem{}Southworth J., Maxted P.~F.~L., Smalley B., 2004a, \mnras, 349, 547 

\bibitem{}Southworth J., Maxted P.~F.~L., Smalley B., 2004b, \mnras, 351, 1277 

\bibitem{}Strassmeier K.~G. et al., 2010, Advances in Astronomy, 19  

\bibitem{}Turner J.~D. et al., 2013, \mnras, 428, 678 

\end{thebibliography}
\end{document}